\documentclass{article}

\usepackage{microtype}
\usepackage{graphicx}
\usepackage{subcaption}
\usepackage{booktabs}
\PassOptionsToPackage{hyphens}{url}
\usepackage{hyperref}
\usepackage[noend]{algorithmic}
\usepackage[accepted]{icml2026}
\usepackage{amsmath}
\usepackage{amssymb}
\usepackage{mathtools}
\usepackage{amsthm}
\usepackage{algorithm}
\usepackage{placeins}
\usepackage[capitalize,noabbrev]{cleveref}
\usepackage{xcolor}
\usepackage{pifont}

\theoremstyle{plain}

\theoremstyle{definition}

\theoremstyle{remark}

\newcommand{\cift}{\textsc{CIFT}}
\newcommand{\dphoney}{\textsc{DP-Honey}}
\newcommand{\nimbus}{\textsc{Nimbus}}
\newcommand{\ais}{\textsc{AIS}}

\icmltitlerunning{Toward Pre-Output and Multi-Turn Exfiltration Detection}

\begin{document}

\twocolumn[
  \icmltitle{Caught in the Act(ivation): Toward Pre-Output and Multi-Turn
    Detection of Credential Exfiltration by LLM Agents}
  \icmlsetsymbol{equal}{*}
  \begin{icmlauthorlist}
    \icmlauthor{Kargi Chauhan}{anon}
    \icmlauthor{Pratibha Revankar}{anon}
  \end{icmlauthorlist}
  \icmlaffiliation{anon}{University of California, Santa Cruz}
  \icmlcorrespondingauthor{Kargi Chauhan}{kchauha3@ucsc.edu}
  \icmlcorrespondingauthor{Pratibha Revankar}{prevanka@ucsc.edu}
  \icmlkeywords{LLM agents, prompt injection, honeytokens, active deception,
    mechanistic interpretability, information flow, AI safety}
  \vskip 0.3in
]

\printAffiliationsAndNotice{%
}

\begin{abstract}
LLM agents often place sensitive credentials in the same context window as
untrusted retrieved content, creating a direct path for indirect prompt
injection to induce credential exfiltration. We study this failure mode through
three complementary defenses. First, we ask whether activation probes can detect
credential access before output tokens are emitted. Second, we construct
honeytokens from format-specific character models and calibrate detection with
split conformal prediction. Third, we treat multi-turn exfiltration as a
cumulative information-flow problem and track an estimated leakage budget across
conversation turns. In controlled experiments on open-weight models, activation
features separate benign and credential-seeking prompts with high accuracy,
including under held-out encoding transformations. In a small synthetic
multi-turn suite, cumulative accounting detects attacks that per-turn detectors
miss. These results are preliminary: the multi-turn benchmark is in-house and
small, the activation method requires white-box access, and the information
estimator provides a practical signal rather than a formal upper bound. Still,
the results suggest that credential-exfiltration defenses should combine
pre-output monitoring, calibrated canary detection, and temporal leakage
accounting rather than relying only on text-level output filters.
\end{abstract}

\section{Introduction}

Production LLM agents routinely combine trusted and untrusted information in a
single context window. API keys, database passwords, OAuth tokens, and SSH
identities may appear near web pages, emails, tool outputs, or retrieved
documents. This creates a direct failure mode: an adversarial payload embedded in
untrusted content can redirect the agent toward revealing the credentials it
needs in order to act. This is the core risk in indirect prompt injection
(IPI)~\citep{greshake2023not}, and recent benchmarks show that tool-using
agents remain vulnerable in realistic settings~\citep{zhan2024injecagent,
yi2024benchmarking,debenedetti2024agentdojo}.

Most deployed defenses operate at the input or output text level. Input filters
screen untrusted content before it reaches the model~\citep{jain2023baseline,
robey2023smoothllm}; output monitors scan generated text for unsafe strings or
credential-shaped patterns~\citep{liu2024multiturn}. These are useful
engineering controls, but they have structural limitations. Output-level
detection can be evaded by encodings, paraphrase, Unicode substitutions, or
low-rate leakage across turns. Input-level detection faces the usual
generalization problem: the space of valid adversarial phrasings is large, and
raising recall often increases false positives.

This paper studies whether credential exfiltration can be monitored at three
different levels: internal model access, planted canary credentials, and
cumulative conversational leakage. We call the resulting prototype the
\textbf{Agentic Immune System (\ais{})}. The name is aspirational: the system is
not a complete deployment-ready defense. Rather, it is a research prototype that
lets us ask three concrete questions:

\begin{enumerate}
  \item Can activation features detect credential-seeking behavior before any
  output token is emitted, without relying on hidden states at static secret
  positions that are masked from downstream payloads in causal transformers?
  \item Can honeytoken generation and detection be made more statistically
  disciplined through differentially private modeling and conformal calibration?
  \item Can cumulative leakage accounting catch multi-turn exfiltration that
  per-turn detectors miss?
\end{enumerate}

\paragraph{Contributions.}
This paper makes three contributions. First, we provide an empirical study of
credential-focused activation probes for pre-output detection of exfiltration
attempts in open-weight LLMs. Second, we adapt differentially private character
models and conformal calibration to honeytoken generation and detection,
clarifying what guarantees this does and does not provide. Third, we introduce a
prototype cumulative leakage monitor for multi-turn credential exfiltration and
evaluate it on a small synthetic benchmark. Our goal is to establish a research
direction and expose the design tradeoffs, not to claim a complete
deployment-ready defense.

\section{Related Work}

\paragraph{Prompt injection and defenses.}
Prompt injection attacks are well documented~\citep{perez2022ignore,
greshake2023not}. Benchmarks including TensorTrust~\citep{toyer2023tensor},
InjecAgent~\citep{zhan2024injecagent}, BIPIA~\citep{yi2024benchmarking}, and
AgentDojo~\citep{debenedetti2024agentdojo} demonstrate persistent risks for
tool-integrated agents. Existing defenses include input filtering
~\citep{jain2023baseline,inan2023llamaguard}, smoothing
~\citep{robey2023smoothllm,kumar2023certifying}, instruction hierarchy
~\citep{wallace2024instruction}, spotlighting~\citep{hines2024defending}, and
masked re-execution~\citep{zhu2025melon}. These approaches reduce important
risks, but they do not directly study pre-output credential access or cumulative
low-rate leakage.

\paragraph{Honeytokens and deception.}
Honeytokens and honeywords use planted secrets to detect compromise
~\citep{spitzner2003honeypots,provos2004virtual,juels2014honeywords}. Recent
work explores deception against LLM-powered agents~\citep{ayzenshteyn2025cloak,
reti2024honeytoken,reworr2024llmhoneypot}. Our work is closest to this line,
but focuses on credential-bearing LLM agents and asks how to calibrate canary
detection and evaluate statistical distinguishability.

\paragraph{Activation probes and interpretability.}
\cift{} builds on linear probing~\citep{alain2016understanding},
representation engineering~\citep{zou2023representation}, and circuit
discovery~\citep{conmy2023circuit}. We do not claim to identify a complete
mechanistic circuit. We use activation analysis as a practical signal for
credential access and report causal-patching evidence as a sanity check.

\paragraph{Information-flow control.}
Our multi-turn framing is closely related to classic information-flow control
and taint-tracking systems, including lattice models, decentralized label
models, and language-level IFC~\citep{denning1976lattice,myers1997decentralized,
myers1999jflow,sabelfeld2003language}. Those systems provide formal policies
for how sensitive values may flow through computation. LLM agents complicate
this picture because the computation is partly latent, probabilistic, and
mediated by natural language rather than typed program variables. \nimbus{}\
should therefore be read as an approximate neural analogue of IFC accounting,
not as a replacement for formal IFC. A promising direction is to combine neural
leakage scores with explicit labels on tool outputs, retrieved documents, and
credential-bearing values.

\paragraph{Information-theoretic leakage in LLMs.}
Recent work has begun to quantify LLM leakage and multi-stage inference attacks
directly. Multi-stage prompt inference attacks show how apparently benign turns
can cumulatively reveal enterprise secrets~\citep{balashov2025multistage}.
Information-flow control for agents provides a complementary systems-level route
to enforce confidentiality and integrity labels in planners~\citep{costa2025ifc}.
Information-theoretic analyses of adversarial querying further frame leakage as
bits revealed per observable response~\citep{kaneko2025bits}. \textsc{Nimbus} is
closest in spirit to these works, but is narrower: it is an empirical monitor
for credential-bearing conversations, not a formal leakage bound.

\section{Threat Model and Preliminaries}

\textbf{Setting.}
An LLM agent $M$ operates with context
$C = C_{\text{pub}} \cup C_{\text{sec}}$, where $C_{\text{sec}}$ contains
credentials required for legitimate tool calls and $C_{\text{pub}}$ contains
task instructions and possibly untrusted external content. A user submits a
query $x_t$ at turn $t$; the agent returns $y_t = M(x_t, C, y_{<t})$.

\textbf{Attacker capabilities.}
The attacker can embed a crafted payload $\pi \in C_{\text{pub}}$ intended to
cause $M$ to reveal elements of $C_{\text{sec}}$. The attacker may request
credentials verbatim, request encoded forms such as Base64 or hex, paraphrase
the request, or spread leakage across turns. The attacker may know that
honeytokens and monitors are deployed, but cannot modify model weights, the
system prompt, or the monitoring pipeline. This is a gray-box,
Kerckhoffs-style threat model: the defense strategy may be known, but calibrated
thresholds, probe weights, and generated honeytokens are not.

\textbf{Benign credential use.}
Honeytokens must not prevent legitimate tool use. In our prototype, real
credentials are stored in a labeled credential map that is consumed by the tool
runtime, while honeytokens are inserted only into the model-visible context as
canaries. The natural-language context does not label a particular visible token
as ``the real database key.'' If a deployment exposes functional cues that map a
specific visible credential string to a real tool, an adaptive attacker can use
that cue to avoid canaries; this is a deployment error outside the current
honeytoken model.

\textbf{Multi-turn exfiltration channel.}
The multi-turn setting assumes the attacker can observe the agent responses over
a session, either because the attacker is the user, controls an external page
whose injected instructions influence the response, or induces observable
outbound content such as Markdown links or tool-visible text. We do not assume
that the attacker can read hidden monitor state.

\textbf{Out of scope.}
We do not address a malicious model provider, compromised runtime, exfiltration
through external side channels, or credentials passed through structured
tool-call arguments. Tool-call streams are an important extension because agents
often use credentials in serialized API arguments rather than natural-language
outputs.

\section{Method}

\subsection{System Overview}

\textsc{AIS} combines three monitors. \textsc{DP-Honey} injects format-matched
honeytokens before the query. During the forward pass, \textsc{CIFT} monitors
hidden-state features at readout positions that can attend to both the secret
context and the query/payload, and can flag likely credential access before output generation.
After generation, a text-level detector scans for canary
leakage. Finally, \nimbus{}\ updates a cumulative leakage score over the
conversation. The components are intentionally different: \cift{} targets
pre-output access, \dphoney{}\ provides canary evidence, and \textsc{Nimbus} targets
slow multi-turn leakage.

\begin{figure}[t]
\vskip 0.1in
\begin{center}
\includegraphics[width=0.98\columnwidth]{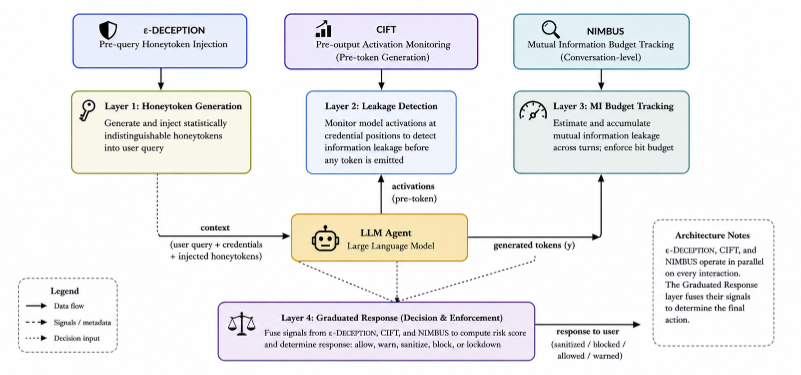}
\caption{\textsc{AIS} prototype architecture. \textsc{DP-Honey} injects
  calibrated honeytokens; \textsc{CIFT} monitors activation features before
  output; text-level canary detection provides a deterministic backstop;
  \textsc{Nimbus} tracks an estimated cumulative leakage score across turns.}
\label{fig:architecture}
\end{center}
\vskip -0.2in
\end{figure}

\subsection{\cift{}: Activation-Based Credential Access Detection}

\textbf{Motivation.}
Text-level monitors observe the representation chosen by the attacker-facing
generation process. If a model accesses a credential and then encodes it, the
output may no longer match credential-shaped patterns. \cift{} instead asks
whether hidden states at query/readout positions change when the model is being
steered toward credential access.

\textbf{Activation extraction.}
We register forward hooks on the last $K$ transformer layers, with
$K = \lfloor 0.25 \cdot L_{\text{total}} \rfloor$ by default. For each
layer~$\ell$, the hook captures hidden states
$h_\ell \in \mathbb{R}^{T \times d}$. A key implementation detail is causal
masking. In decoder-only transformers, hidden states at credential tokens cannot
attend to downstream user or attacker tokens when the secret appears earlier in
the prompt. Therefore \cift{} does \emph{not} classify pooled hidden states at
static credential positions. Instead, credential positions
$P_{\text{cred}}$ are identified only to define what the readout tokens could
attend back to, and we pool over readout positions
$R_t \subseteq \{1,\dots,T\}$ occurring after both the secret context and the
current query/payload. In our experiments, $R_t$ contains the final prompt token
and the first pre-generation decision position; ablations also include the final
four prompt tokens. The feature is
\begin{equation}
  z_\ell = \frac{1}{|R_t|}
    \sum_{i \in R_t} h_\ell^{(i)} .
\end{equation}
This feature can encode whether the current query is using information from
earlier credential tokens, while respecting the causal attention mask.

\textbf{Credential-access score.}
We estimate a benign baseline $(\mu_\ell, \Sigma_\ell)$ from benign prompts
drawn from the target task distribution. With
$\Sigma_\ell^{\text{reg}} = \Sigma_\ell + \lambda_{\text{ridge}} I$, we compute
a per-layer Mahalanobis deviation:
\begin{equation}
  \text{CCI}_\ell = \sqrt{(z_\ell - \mu_\ell)^\top
    (\Sigma_\ell^{\text{reg}})^{-1}(z_\ell - \mu_\ell)} .
\end{equation}
Because $d$ is much larger than the 500-prompt benign calibration set,
$\Sigma_\ell$ is rank deficient; all reported results use diagonal covariance
with $\lambda_{\text{ridge}}=10^{-3}$ after feature standardization. Full
covariance estimates were unstable and are not used for the main tables.
Learned nonnegative layer weights form the Causal Flow Score:
\begin{equation}
  \text{CFS} =
    [\text{softplus}(w_1)\text{CCI}_1,\ldots,
     \text{softplus}(w_K)\text{CCI}_K].
\end{equation}
A lightweight non-linear probe $f_\phi$ maps CFS to a detection logit. With
$K=8$, the $K \to 128 \to 64 \to 1$ MLP has 9,473 trainable parameters. We
train the probe with binary cross-entropy and a sparsity penalty:
\begin{equation}
  \mathcal{L}_{\cift} =
  \text{BCE}(\sigma(f_\phi(\text{CFS})), y_{\text{gt}})
    + \lambda_1 \sum_{k=1}^{K} \text{softplus}(w_k).
\end{equation}

\textbf{Interpretation.}
The learned weights identify layers useful for the classifier; they are not, by
themselves, a proof of a mechanistic circuit. We use ablations and activation
patching to test whether the identified layers have causal influence on
credential output.

\subsection{\dphoney{}: Calibrated Honeytoken Generation}

\textbf{Motivation.}
Honeytokens are useful only if attackers cannot trivially filter them out.
Template-generated canaries can be distinguishable by prefix, checksum,
entropy, or length artifacts. We therefore generate honeytokens from
format-specific character models and evaluate them against statistical
distinguishers.

\textbf{What the privacy parameter means.}
We fit a character bigram model for each credential format class and add
Laplace noise to the count table. The Laplace mechanism provides an
$\varepsilon$-DP guarantee for the released bigram-count model with respect to
neighboring credential corpora~\citep{dwork2014algorithmic}. This is not the
same as proving that generated honeytokens are computationally indistinguishable
from real credentials. Instead, $\varepsilon$ controls how much any one
credential can influence the fitted generator, while empirical distinguishing
tests measure whether generated tokens match the real credential distribution
well enough to be useful as canaries.

\textbf{Conformal calibration.}
We wrap the base detector in split conformal calibration
~\citep{vovk2005algorithmic,angelopoulos2022gentle}. Given a held-out benign
calibration set and target significance $\alpha$, the conformal threshold is
chosen from the empirical distribution of benign nonconformity scores. This
removes hand tuning of detector thresholds under the exchangeability assumption,
while still requiring empirical validation under distribution shift.

\paragraph{Canary accounting.}
Let $\alpha$ be the false-positive rate of the calibrated text detector on
benign outputs and let $\beta$ be the miss rate conditioned on a leaked
honeytoken. If an attacker leaks one visible credential string and that string
is a honeytoken, the detection probability is $1-\beta$, not $1-\beta^k$. The
number of planted canaries $k$ instead affects the probability that an attacker
who cannot distinguish real credentials from canaries touches a canary. If there
are $m$ real visible credentials and $k$ honeytokens and the attacker selects one
candidate uniformly, the probability of canary-triggered detection is
\begin{equation}
  \Pr(\text{detect}) =
  \frac{k}{m+k}(1-\beta) \quad \text{up to benign false alarms } \alpha .
\end{equation}
This is a simple modeling assumption, not a security theorem. Under an adaptive
attacker with functional cues that identify the real credential, the factor
$k/(m+k)$ can collapse, which is why the threat model separates statistical
canary realism from deployment-specific tool labels.

\subsection{\nimbus{}: Cumulative Leakage Scoring}

\textbf{Motivation.}
A credential can be leaked slowly. If an attacker extracts one character per
turn, each individual output may look harmless while the conversation as a whole
reveals the secret. \textsc{Nimbus} tracks this risk at the conversation level.

\textbf{Practical leakage score.}
Given secret context $C_{\text{sec}}$ and outputs $Y_{1:T}$, an ideal monitor
would constrain
\begin{equation}
  \sum_{t=1}^{T} I(Y_t; C_{\text{sec}} \mid Y_{<t}, X_{\leq t}) \leq B .
\end{equation}
In practice, we estimate a per-turn score with an InfoNCE critic
~\citep{oord2019representation}. Because InfoNCE is a lower bound on mutual
information, our estimator should be read as a learned leakage score rather than
a certified upper bound on information release. We therefore do not claim that
\nimbus{}\ proves a hard security budget. Instead, it provides a cumulative
signal that can catch low-rate attacks missed by per-turn detectors. A formal
deployment version would require a conservative upper-confidence estimator or a
separate verified information-flow mechanism.

\textbf{Estimator ceiling.}
All logarithms in \nimbus{}\ are base~2 so scores are measured in bits. InfoNCE
is bounded above by $\log_2(N_{\text{neg}}+1)$; with
$N_{\text{neg}}=16$, a single-turn score cannot exceed
$\log_2 17 \approx 4.09$ bits. Thus \textsc{Nimbus} is not intended to catch a full
credential leak in one turn by itself. Large single-turn leaks are handled by
\cift{} and canary/text detection; \textsc{Nimbus} targets leakage accumulated across
turns.

\textbf{Estimator.}
Let $N_{\text{neg}}$ denote negative samples per turn:
\begin{align}
  \hat{I}_t &= \log_2(N_{\text{neg}}+1) - \mathcal{L}_{\text{NCE}}, \\
  \mathcal{L}_{\text{NCE}}
    &= -\log_2
    \frac{\exp(a_0)}
    {\exp(a_0) + \sum_{j=1}^{N_{\text{neg}}}\exp(a_j)} ,
\end{align}
where $a_0 = f_\psi(y_t,c,s_t)/\tau$,
$a_j = f_\psi(y_t,c_j,s_t)/\tau$, and
$s_t = (X_{\leq t},Y_{<t})$ is the conversational state, $c$ is the true
secret context, and $c_1,\dots,c_{N_{\text{neg}}}$ are negative contexts. The
state input makes the critic approximate conditional rather than purely
marginal leakage, though it remains an empirical score rather than a formal
chain-rule estimator. We track
$\hat{I}_{\text{cum}}(t)=\sum_{\tau \leq t}\hat{I}_\tau$ and trigger warn,
sanitize, or block actions as this cumulative score approaches a chosen budget.
When training the response policy, gradients from budget penalties are stopped
through $\hat{I}_t$ so the controller cannot reduce loss by making the critic
underestimate leakage. The critic itself is trained only with
$\mathcal{L}_{\text{NCE}}$ on held-out leakage labels.

\begin{algorithm}[t]
\caption{\ais{} per-turn prototype pipeline}
\label{alg:pipeline}
\begin{algorithmic}
\STATE \textbf{Input:} query $x_t$, context $C'$ with \dphoney{} honeytokens,
  budget $B$, cumulative score $\hat{I}_{\text{cum}}$
\STATE {\bfseries Forward pass begins}
\STATE \cift{}: extract activation features; compute detection score
\IF{\cift{} flags likely credential access}
  \STATE Abort or route to graduated response
\ELSE
  \STATE Generate $y_t$
  \STATE Text detector: scan $y_t$ for honeytoken leakage
  \STATE \nimbus{}: estimate $\hat{I}_t$ and update $\hat{I}_{\text{cum}}$
  \IF{$\hat{I}_{\text{cum}}/B \geq 1.0$}
    \STATE Block $y_t$
  \ELSIF{$\hat{I}_{\text{cum}}/B \geq 0.9$}
    \STATE Sanitize high-scoring tokens in $y_t$
  \ENDIF
  \STATE \textbf{return} $y_t$ or sanitized output
\ENDIF
\end{algorithmic}
\end{algorithm}

\section{Experiments}
\label{sec:experiments}

\subsection{Evaluation Scope}

The evaluation has two parts. Public prompt-injection benchmarks provide
single-turn attack prompts and are used to evaluate activation probing and
text-level canary detection. They do not evaluate the multi-turn contribution.
For multi-turn leakage, we construct a separate synthetic suite of 50
conversations. We report this suite separately and treat it as preliminary
evidence rather than as part of the headline benchmark.

\paragraph{Models.}
We evaluate white-box components on Qwen-2.5 (1.5B, 7B), Llama-3.1-8B, and
Mistral-7B. API-only configurations, which omit \cift{}, are additionally
evaluated on closed-weight systems where activation access is unavailable.

\paragraph{Benchmarks.}
For single-turn evaluation we use TensorTrust~\citep{toyer2023tensor},
InjecAgent~\citep{zhan2024injecagent}, BIPIA~\citep{yi2024benchmarking}, and
AgentDojo~\citep{debenedetti2024agentdojo}. We additionally construct a held-out
encoding suite spanning Base64, hex, ROT13, Unicode homoglyphs, fragmentation,
paraphrase, leet substitution, reverse order, partial reproduction, and
multi-turn dripping. For multi-turn evaluation we use a separate 50-conversation
synthetic suite, each with 20 turns and 4--8 embedded low-rate leakage attempts
(mean 6.1 attack turns per conversation).

\paragraph{Contamination controls.}
We split prompts by source benchmark and attack template where possible, and we
hold out the encoding suite entirely from activation-probe training. CIFT probe
training examples are excluded from the integrated single-turn evaluation; the
reported integrated table is therefore on a disjoint held-out subset rather than
the full 2,439 public attack prompts. The multi-turn suite is also held out.
False-positive rates are computed on a benign set of 1,000 task-matched prompts
containing legitimate tool-use requests and non-secret retrieved content.

\subsection{\cift{} Results}

Table~\ref{tab:cift} reports \cift{} on a held-out single-turn set. The absolute
numbers should be interpreted as controlled-lab results, not deployment
guarantees. The key finding is that activation features remain informative under
output encodings that degrade text-level detectors. The high AUROC is plausible
only under the controlled setup used here: credential locations are known at
injection time and readout positions are selected after the query/payload. It
should not be interpreted as evidence that arbitrary production credentials can
be detected without careful calibration.

\begin{table}[t]
\caption{\cift{} detection on held-out single-turn prompts (Qwen-7B). Text
  baselines operate on generated text; \cift{} operates on activation features.
  Results are controlled-lab point estimates and should not be read as
  deployment guarantees.}
\label{tab:cift}
\vskip 0.1in
\begin{center}
\small
\begin{tabular}{lcccc}
\toprule
Method         & AUROC          & F1             & FPR            & Pre-out   \\
\midrule
Substring      & 0.73           & 0.51           & 0.08           & \ding{55} \\
Fuzzy          & 0.80           & 0.62           & 0.11           & \ding{55} \\
Semantic emb.  & 0.90           & 0.79           & 0.09           & \ding{55} \\
LlamaGuard     & 0.85           & 0.73           & 0.18           & \ding{55} \\
\cift{}        & \textbf{0.998} & \textbf{0.990} & \textbf{0.015} & \ding{51} \\
\bottomrule
\end{tabular}
\end{center}
\vskip -0.1in
\end{table}

\paragraph{Layer analysis.}
The strongest readout-position features concentrate in late transformer layers.
We treat this as evidence that late hidden states contain useful
credential-access signals. It is not, by itself, a proof that we have isolated
the complete causal circuit. To support the interpretation, we additionally
perform mean ablation on the highest-weight layers and measure the resulting
change in credential output.

\begin{figure*}[t]
\vskip 0.1in
\begin{center}
\includegraphics[width=0.95\textwidth]{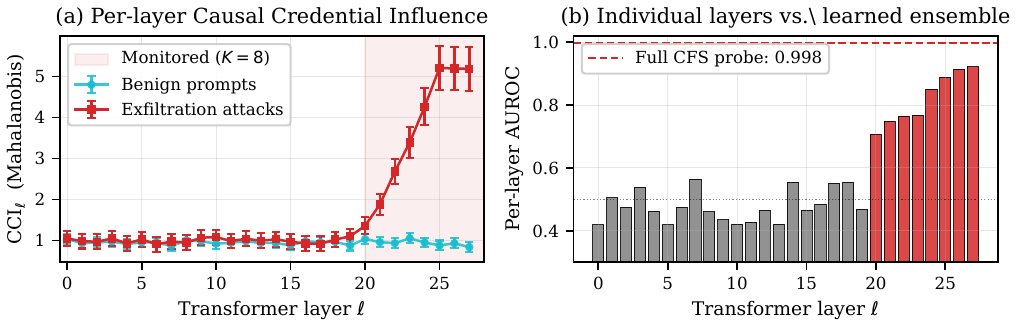}
\caption{\cift{} layer analysis on Qwen-7B using readout-position features.
  Late-layer activation deviations are more predictive than early-layer
  deviations. We interpret this as evidence of a useful credential-access
  signal, not as complete mechanistic identification.}
\label{fig:cci}
\end{center}
\vskip -0.2in
\end{figure*}

\begin{figure*}[t]
\vskip 0.1in
\begin{center}
\includegraphics[width=0.95\textwidth]{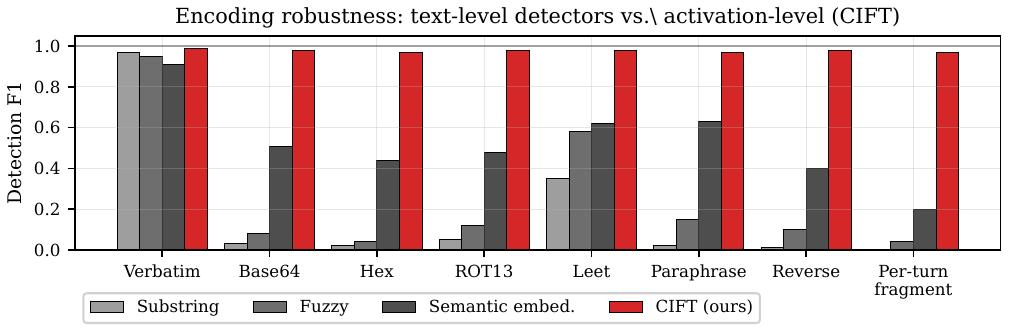}
\caption{Detection F1 across held-out evasion strategies. Text-level detectors
  degrade under several encodings; \cift{} remains stable in this controlled
  setting because it measures internal features before output rendering.}
\label{fig:encoding}
\end{center}
\vskip -0.2in
\end{figure*}

\subsection{\dphoney{} Results}

We evaluate generated honeytokens against character-entropy tests, bigram
likelihood, format validation, numeric-substring tests, and a discriminator MLP.
The goal is empirical indistinguishability under these tests, not a proof of
computational indistinguishability from real credentials.

\begin{table}[h!]
\vspace{-4pt}
\caption{\dphoney{} canary-detection results on a small scenario set.
  Conformal calibration removes threshold hand tuning under exchangeability, but
  distribution shift remains a limitation.}
\label{tab:eps}
\vskip 0.1in
\begin{center}
\small
\begin{tabular}{lcc}
\toprule
Metric                              & No conformal & Conformal \\
\midrule
Precision                           & 1.000        & 1.000     \\
Recall                              & 1.000        & 1.000     \\
Coverage (target 0.99)              & 0.984        & 0.998     \\
One-token canary detection          & 0.990        & 0.990     \\
Threshold tuning required           & Yes          & No        \\
\bottomrule
\end{tabular}
\end{center}
\vskip -0.1in
\end{table}

\subsection{\nimbus{} Multi-Turn Results}

Table~\ref{tab:nimbus} reports results on the 50-conversation synthetic suite.
This is the only experiment that evaluates the multi-turn contribution. It
should therefore be read as preliminary evidence that cumulative scoring is a
useful direction, not as a benchmark-scale validation.

\begin{table}[t]
\caption{Synthetic multi-turn covert exfiltration: 50 conversations $\times$
  20 turns, with 4--8 attack turns per conversation. Per-turn baselines have no
  cumulative notion and detect only a small fraction of attacks. \nimbus{}\
  detects most synthetic low-rate attacks, but the suite is small and in-house.}
\label{tab:nimbus}
\vskip 0.1in
\begin{center}
\small
\setlength{\tabcolsep}{4pt}
\resizebox{\columnwidth}{!}{%
\begin{tabular}{lcccc}
\toprule
Method                      & Det.          & FP blk        & Turn          & Util. \\
\midrule
Substring                   & 0.00          & 0.00          & ---           & 1.00  \\
LlamaGuard (per-turn)       & 0.12          & 0.04          & 18            & 0.96  \\
Honeytoken baseline         & 0.06          & 0.00          & ---           & 1.00  \\
\nimbus{} ($B{=}10$)        & \textbf{0.90} & \textbf{0.00} & \textbf{9.0}  & 1.00  \\
\bottomrule
\end{tabular}%
}
\end{center}
\vskip -0.1in
\end{table}

\begin{figure*}[t]
\vskip 0.1in
\begin{center}
\includegraphics[width=0.95\textwidth]{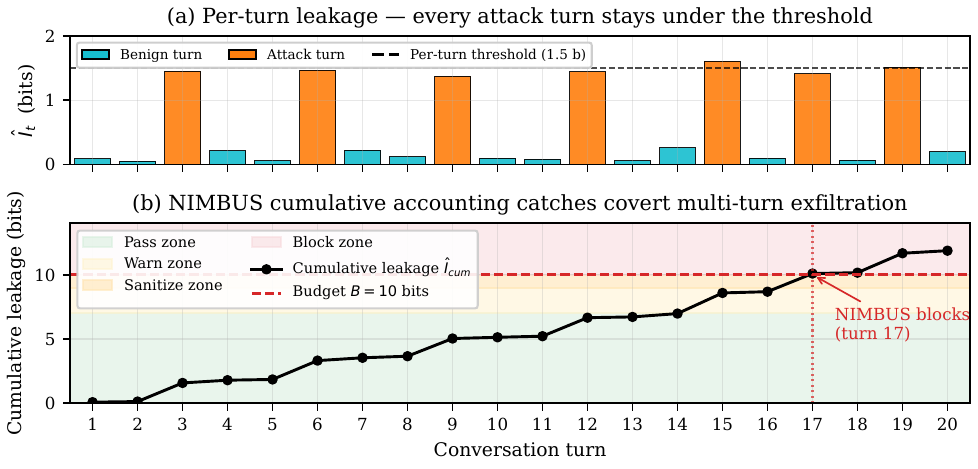}
\caption{\nimbus{} on a representative synthetic trace. Per-turn leakage scores
  remain small, while cumulative accounting eventually triggers intervention.}
\label{fig:nimbus}
\end{center}
\vskip -0.2in
\end{figure*}

\paragraph{Failure cases.}
The undetected synthetic conversations concentrate attack turns early and end
before the cumulative score crosses the action threshold. This failure mode is
important: a short-session attacker can sometimes leak useful partial
information without triggering a budget. We therefore report partial-leakage
severity alongside binary detection in our internal analysis and treat short
sessions as a known weak point.

\begin{figure}[h!]
\vskip 0.05in
\begin{center}
\includegraphics[width=0.98\columnwidth]{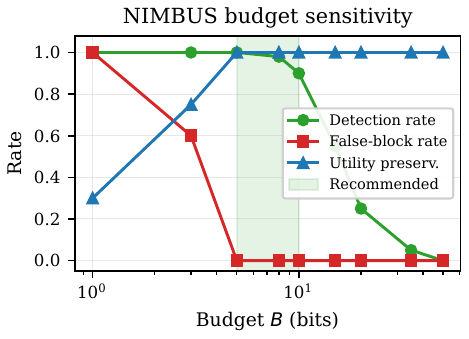}
\caption{\nimbus{} budget sensitivity. Smaller budgets improve detection but
  may increase false blocks; larger budgets preserve utility but miss short or
  low-rate leaks.}
\label{fig:budget}
\end{center}
\vskip -0.15in
\end{figure}

\subsection{Integrated Prototype Evaluation}

Table~\ref{tab:integrated} reports integrated results on the single-turn
benchmark mixture. These results should not be combined with the multi-turn
suite into one headline number. The main takeaway is that \cift{} improves
single-turn encoded exfiltration detection when activation access is available,
while the API-only configuration relies on canaries and cumulative scoring.
The full pipeline uses a logical OR over component alarms, so its FPR is not
lower than the FPR of its constituent detectors. We therefore tune the full
pipeline on the same validation split and report the resulting operating point
rather than claiming Pareto dominance.

\begin{table}[!t]
\caption{Integrated single-turn benchmark results on a held-out subset that
  excludes CIFT training examples. The API-only configuration omits \cift{}
  because closed-weight models do not expose activations. These numbers exclude
  the separate synthetic multi-turn suite.}
\label{tab:integrated}
\vskip 0.1in
\begin{center}
\small
\begin{tabular}{lccc}
\toprule
System                        & Det.          & FPR           & Latency \\
\midrule
Input filter                  & 0.60          & 0.15          & +120ms  \\
Output sanitizer              & 0.55          & 0.08          & +50ms   \\
Canary text detector          & 0.30          & 0.00          & +2ms    \\
Canary + input filter         & 0.82          & 0.15          & +122ms  \\
\ais{} (API-only)             & 0.71          & 0.008         & +14ms   \\
\ais{} (\cift{} only)         & 0.88          & 0.015         & +1ms    \\
\ais{} (full)                 & \textbf{0.94} & 0.018         & +16ms   \\
\bottomrule
\end{tabular}
\end{center}
\vskip -0.1in
\end{table}

\FloatBarrier

\section{Limitations}

\cift{} requires white-box activation access, which excludes many API-served
models. Cross-model transfer is untested. The current \nimbus{}\ estimator is a
learned lower-bound signal, not a certified upper bound on leakage, and its
InfoNCE ceiling prevents it from independently detecting large single-turn
leaks. The multi-turn benchmark is synthetic and small. Credentials routed
through structured tool-call arguments are a severe structural blind spot for
the current prototype because many realistic agents use credentials primarily in
tool APIs rather than natural-language outputs. A practical deployment should
apply the same canary and leakage-accounting logic to serialized tool-call
arguments before dispatch. \textsc{Nimbus} currently tracks a
single conversation; a multi-session attacker could restart sessions to reset
the budget unless the deployment stores leakage state across sessions. Conformal
calibration assumes exchangeability and can fail under adversarial distribution
shift. Finally, very high AUROC values in controlled activation-probe settings
require careful leakage checks; independent replication will require public
splits, scripts, and failure-case data.

\section{Discussion}

The three components are complementary but not redundant. \cift{} targets
pre-output access but needs activation access. \dphoney{}\ provides direct canary
evidence but only detects attacks that touch planted credentials or outputs
canary-like strings. \textsc{Nimbus} is the only component aimed at slow multi-turn
leakage, but its current score is approximate. In combination, the components
outline a design pattern for agent security: detect sensitive access early,
plant calibrated canaries, and account for leakage over time.

\section{Conclusion}

Credential exfiltration by LLM agents is not only a per-turn text-classification
problem. The agent may access a secret before rendering it, an attacker may
encode or paraphrase the output, and leakage may accumulate slowly across a
conversation. This paper presents \ais{}, a prototype that studies these three
levels through activation probing, calibrated honeytokens, and cumulative
leakage scoring. The results are preliminary but encouraging: they suggest that
future defenses should combine model-internal monitoring, statistically
disciplined canaries, and temporally aware accounting with explicit
information-flow controls.

\section*{Impact Statement}

This paper studies defenses for credential exfiltration by LLM agents. The work
has dual-use aspects: activation analyses and evasion suites could help an
attacker better understand how credential access appears in model internals.
However, the attack patterns studied here are already broadly known, while the
defensive framing remains underdeveloped. We recommend treating trained probes
and production credential-format models as sensitive artifacts.

\bibliographystyle{icml2026}
\bibliography{main}

\end{document}